\documentclass[fleqn,10pt]{wlscirep}
\usepackage[utf8]{inputenc}
\usepackage[T1]{fontenc}
\usepackage{graphicx}
\usepackage{amssymb}
\usepackage{setspace}
\usepackage{booktabs}   
\usepackage{mathtools}
\usepackage{bm, xspace}  
\newcommand{\bmath}[1]{\ensuremath{\bm{#1}}\xspace}
\usepackage{multirow}
\usepackage{tikz}

\newcommand{\bmu}{\bmath{\mu}}

\newcommand{\bz}{\bmath{z}}

\newcommand{\bx}{\bmath{x}}

\DeclareMathOperator*{\argmax}{arg\,max}

\usepackage{algorithmicx}
\usepackage{algorithm}
\usepackage{algpseudocode}
\algnewcommand\algorithmicinput{\textbf{Input:}}
\algnewcommand\Input{\item[\algorithmicinput]}
\algnewcommand\algorithmicinitilize{\textbf{Initialize:}}
\algnewcommand\Initialize{\item[\algorithmicinitilize]}
\algnewcommand\algorithmicoutput{\textbf{Output:}}
\algnewcommand\Output{\item[\algorithmicoutput]}
\algnewcommand\algorithmicalg{\textbf{Algorithm:}}
\algnewcommand\Algorithm{\item[\algorithmicalg]}
\algnewcommand\algorithmicnote{\textbf{Note:}}
\algnewcommand\Note{\item[\algorithmicnote]}

\title{Handling highly correlated genes in prediction analysis of genomic studies}

\author[2]{Li Xing}
\author[1]{Songwan Joun}
\author[1]{Kurt Mackey}
\author[1]{Mary Lesperance}
\author[1, *]{Xuekui Zhang}
\affil[1]{Department of Mathematics and Statistics, University of Victoria, Victoria, Canada}
\affil[2]{Department of Mathematics and Statistics, University of Saskatchewan, Saskatoon, Canada}

\affil[*]{Corresponding author: Xuekui@UVic.ca}

\begin{abstract}
Background: Selecting feature genes to predict phenotypes is one of the typical tasks in analyzing genomics data. Though many general-purpose algorithms were developed for prediction, dealing with highly correlated genes in the prediction model is still not well addressed. High correlation among genes introduces technical problems, such as multi-collinearity issues, leading to unreliable prediction models. Furthermore, when a causal gene (whose variants have an actual biological effect on a phenotype) is highly correlated with other genes, most algorithms select the feature gene from the correlated group in a purely data-driven manner. Since the correlation structure among genes could change substantially when condition changes, the prediction model based on not correctly selected feature genes is unreliable. Therefore, we aim to keep the causal biological signal in the prediction process and build a more robust prediction model.\\
Method: We propose a grouping algorithm, which treats highly correlated genes as a group and uses their common pattern to represent the group's biological signal in feature selection. Our novel grouping algorithm can be integrated into existing prediction algorithms to enhance their prediction performance. Our proposed grouping method has two advantages. First, using the gene group's common patterns makes the prediction more robust and reliable under condition change. Second, it reports whole correlated gene groups as discovered biomarkers for prediction tasks, allowing researchers to conduct follow-up studies to identify causal genes within the identified groups.\\
Result: Using real benchmark scRNA-seq datasets with simulated cell phenotypes, we demonstrate our novel method significantly outperforms standard models in both (1) prediction of cell phenotypes and (2) feature gene selection.
\end{abstract}
\begin{document}

\flushbottom
\maketitle
%
%
\thispagestyle{empty}


\section*{Background}
Association and prediction are two typical tasks in genomic data analysis. Researchers usually use Differential Expression (DE) methods to investigate the association between phenotypes and every 'individual' gene and select feature genes that significantly distinguish the treatment and control groups. Popular DE methods were described in a benchmark study (Soneson and Robinson, 2018). This work focuses on the other typical task, prediction, which involves selecting feature genes and precisely predicting phenotypes. Note that the DE methods are not suitable for prediction tasks. Their approach is to look at genes one by one. However, complex phenotypes are often affected by multiple genes jointly. We need to utilize information from multiple genes simultaneously in a prediction model. 

Many general-purpose machine learning methods are well-developed for selecting significant predictors to build a good prediction model. For example, Elastic Net \cite{Zou:2005ex}, a supervised machine learning method, has an outstanding prediction performance, which regulates the coefficients of non-relevant predictors to zero and then remains significant ones in the final model. Therefore, it has been applied to genomic data for various tasks, including (1) discovering a diagnostic test of 2-transcript host RNA signatures for discriminating bacterial versus viral infection in febrile children \cite{Herberg:2016fi}, (2)selecting genes and predicting clinical drug response \cite{Geeleher:2014fa}, (3)predicting resistance of HIV drugs from mutation information \cite{Xing:2019ji}, and (4)working as a building block to construct complex models for the analysis of single-cell sequencing data \cite{Dixit:2016fe}. 

Despite the popularity of supervised machine learning methods applied in genomic studies, the challenges raised by highly correlated genes are not well addressed by existing methods. In general, highly correlated predictors introduce technical issues, such as collinearity, making prediction models unstable. Furthermore, in genomic studies, it may cause critical biological problems. For example, suppose we have a group of highly correlated genes. Some of them are in the causal pathway of a phenotype, and their variants carry actual biological effects affecting the outcome. However, the data-driven machine learning tools may select the other non-causal ones in the group due to the high correlation. Since the correlation structure varies when condition changes, such as heterogeneity of cells and change in environmental conditions, the prediction model resulting from directly applying a supervised learner can be unreliable over condition changes. 
To avoid selecting non-causal genes, we propose a grouping method that detects groups of highly correlated genes and uses groups' common patterns in the prediction model instead of just picking one gene from the group. Working with groups has two benefits. First, we report the identified groups to ensure the causal genes are retained, enabling researchers to conduct follow-up studies to identify the causal ones from the reported groups. For example, Mendelian randomization \cite{Gusev:2016ey, Richardson:2020co} can be carried out to identify causal genes by integrating SNPs and gene expression data in the downstream analysis. Alternatively, results from multiple studies can be combined to filter out some non-causal genes, such as those not in the selected groups of all studies. Second, the common pattern of a group of highly correlated genes is more robust and represents the causal signal, which is better than using an arbitrarily selected one. 


The rest of this paper is organized as follows. Section $2$ introduces our novel grouping method and describes how to integrate it into the Elastic Net method. In Section $3$, based on the real-world benchmark single-cell RNA sequencing (scRNA-seq) datasets \cite{Soneson:2018df}, we simulate their cell phenotypes. Using simulated data, we show our novel grouping method significantly improves the performance of the original Elastic Net method in both gene selection and phenotype prediction. Finally, we include our conclusion and discussion in Section $4$.

\section*{Methods}
Let us introduce our grouping method using a hypothetical extreme example and then generalize it to realistic situations. Assume the true relationship of the expression levels of three genes are represented as $\bx_1= -\bx_2= \bx_3$, showing a perfectly correlation (i.e. their correlations are either $1$ or $-1$). Given this relationship, there is a trivial approach to handling those three genes, outperforming all fancy methods. We treat those perfectly correlated predictors as a group, define their common pattern as $\bz=(\bx_1 -\bx_2+ \bx_3)/3$, and use the common pattern to represent the whole group (as a single predictor) in the prediction model. If the representative $\bz$ is selected as an important predictor, all genes within the group will be labelled as `candidate' feature genes. This grouping idea can be extended to more general situations where genes are not perfectly correlated. Specifically, we group highly (positively or negatively) correlated genes and use the common pattern of each gene group as a predictor in the prediction models. Suppose some important gene groups are discovered in the approach. In that case, researchers can conduct a follow-up study (i.e. collecting additional data) to identify the causal genes within those groups and then use the causal genes to replace the gene groups in the final prediction or classification model.

Based on the above description, the critical step is to group the highly correlated genes. In unsupervised machine learning, clustering methods are designed to group similar objectives. Agglomerative hierarchical clustering and K-Means clustering are both popular methods for such tasks. However, none of them can be applied directly to our specific grouping task. In human genome, there are about $20,000$ protein-coding genes, $25,000$ non-protein-coding genes, and $2300$ micro-RNA genes. The K-Means algorithm is well-known for its ability to handle such large genomic data. However, we cannot apply it directly to solve our problem since it has three issues. First, the K-Means algorithm cannot directly control the magnitude of within-cluster correlations, while our goal is to group only highly correlated genes. Second, our grouping task is required to cluster highly correlated genes, which leads to a large number of clusters. The K-Means algorithm can be significantly slower when the number of clusters is large. Third, the K-Means algorithm is defined based on the Euclidean distance, while we need a method based on correlation. For example, if two strongly related genes have a large negative correlation, $cor(\bx_1, \bx_2) \approx -1$, their Euclidean distance can be very big, but we should group those two genes together and use $(\bx_1 - \bx_2)/2$ as their common pattern. On the other side, in a hierarchical clustering algorithm, the resulting clusters are obtained by cutting a dendrogram. The cutting threshold directly controls the magnitude of the within-cluster correlations of genes, which addresses the first problem perfectly. Still, the hierarchical clustering algorithm cannot handle such a large number of genes. Therefore, we designed a hybrid approach using the divide-and-conquer strategy to combine the strength from both methods. We first pre-group the genes into smaller subsets using a modified K-Means algorithm and then apply the modified hierarchical clustering algorithm to group genes in each subset. Next, we discuss the detailed two steps of our novel grouping method.

\subsection*{Pre-grouping genes by the modified K-Means algorithm}

The standard K-Means algorithm consists of three components: (1) specify the initial cluster membership of each gene; (2) for every gene, the A-step calculates its Euclidean distances between every cluster center and updates its cluster membership to the nearest cluster center; and (3) based on the current cluster memberships of all genes, the U-step updates the mean of the expression profiles of all genes in each cluster and uses those means as the current cluster centers. The final results of K-Means are obtained by iteratively applies A-steps and U-steps until convergence.

We modified the K-Means algorithm with two adjustments to address the third issue with K-Means in our problem. In the A-step, we replace the Euclidean distance by the dissimilarity defined as $1-|\mbox{correlation}|$ to ensure highly correlated genes have small distances to each other. In the U-step, we reverse the sign of expression levels of the genes that are negatively correlated with their current cluster centers before updating the centers. Note, we keep track of genes’ sign-change in every iteration to know the direction of each gene in their final clusters. The detailed steps of our revised algorithm are provided in Algorithm~\ref{alg1}. A good initial clustering is critical for fast convergence. Currently, we use the results of the standard K-Means algorithm as initial clusters for our modified K-Means. 

When applying our modified K-Means to analyze data, we must ensure that all subsets obtained from pre-grouping are small enough to be processed by the hierarchical clustering algorithm. Increasing the number of clusters in the K-Means algorithm does not always lead to splitting the largest subset. Furthermore, as discussed in the second issue above, using a large number of clusters can substantially slow down K-Means. So, we propose to apply the K-Means algorithm iteratively. More specifically, we first apply the K-Means algorithm using $K=10$ clusters. Then we keep on applying the K-Means algorithm to split the largest cluster into $K=10$ sub-clusters until the largest one is small enough (e.g. with less than $1000$ genes). This iterative approach provides an efficient way to limit the size of the largest cluster and solves the second issue discussed above.

\begin{algorithm}
	\caption{Modified K-Means algorithm for pre-clustering genes}
	\label{alg1}
	\begin{algorithmic}
		\Input \\
		(1) Gene expression level: $\boldsymbol{x} = (\boldsymbol{x}_1^T,\dots,\boldsymbol{x}_p^T)$,  a $n \times p$ matrix\\
		(2) Number of clusters: $K$ \\
		Note: $n$ is the number of cells and $p$ is the number of genes. \\
		
		\Initialize \\
		(1) [Cluster memberships: $\boldsymbol{\mathcal{I}}= (\mathcal{I}_1, \ldots, \mathcal{I}_p)$]  Use the results of standard KMeans algorithm with dissimilarity distance to initialize $\mathcal{I}_i=k$, which represents the $i^{th}$ gene belonging to the $k^{th}$ cluster. \\
		(2) [Sign of correlations to its cluster center: $\boldsymbol{S}= (S_1, \ldots, S_p)$] Initialize $S_i=+1$.\\ 
		\Algorithm
		\Repeat
		\For{$k = 1$ to $K$}
		\State{$\bmu_{k} \gets \frac{1}{n_k} \sum_{\{i:\mathcal{I}_i =k\}} S_i\boldsymbol{x}_i$} (Update cluster centers by signed average, $n_k$ is the number of samples in the $k^{th}$ cluster )
		\EndFor
		\For{$i = 1$ to $p$}
		\State{$r_{i,k} \gets cor(\boldsymbol{\mu}_{k}, \boldsymbol{x}_i)$} (Calculate correlation for $k=1,\ldots,K$)
		\State{$\mathcal{I}_i \gets l= \argmax_k |r_{i,k}|$ (Assign $\boldsymbol{x}_i$ into most correlated cluster $l$) }
		\State{$S_i \gets \mbox{sign}(r_{i,l}) $} (Update the sign of correlations to its cluster center)
		\EndFor
		\Until{cluster membership $\boldsymbol{\mathcal{I}}$ remains unchanged across iterations}\\
		
		\Output \\
		$\boldsymbol{\mathcal{I}}$ (cluster membership)  and  $\boldsymbol{S}$ (sign of correlations to its cluster center)
	\end{algorithmic}
\end{algorithm}

\subsection*{Grouping genes by the hierarchical clustering algorithm}
The pre-clustering step splits genes into subsets, which are small enough to be directly handled by a standard hierarchical clustering algorithm.  Note that the Algorithm~\ref{alg1} flips the sign of genes that are negatively correlated with others in the same cluster. Then genes in pre-clustered subsets are all positively correlated. Therefore, we propose parallelly applying a standard hierarchical clustering algorithm to further group genes within each subset. 

In the hierarchical clustering, similar to the pre-clustering step, we define the dissimilarity between the $j$-th and $i$-th genes as $d(\mbox{gene}_j, \mbox{gene}_i) = 1-cor(\bx_j, \bx_i)$, and the  dissimilarity between the $k$-th and $l$-th clusters as $d(\mbox{cluster}_k, \mbox{cluster}_l) = 1-cor(\bmu_k, \bmu_l)$, where $\bmu_k$ denotes the common pattern of the $k$-th gene-cluster as defined in Algorithm~\ref{alg1}. Since the order of the correlation operation and the average operation are interchangeable, the definition of dissimilarity between clusters is mathematically equivalent to using the average linkage in hierarchical clustering. In summary, we propose to use dissimilarity $1-$ correlation and average linkage in the hierarchical clustering step.

For each subset, the hierarchical clustering algorithm produces a tree-like dendrogram from the bottom (leaves) to the top (root) to represent the correlation structure of genes. The genes closer to the bottom are more correlated. To group highly correlated genes, we trim the trees with a threshold and assign the genes within the same cut branch as one cluster. The trimming threshold controls the strength of the correlation is required to group two genes, so we need a consistent dissimilarity threshold to trim all the trees. There are no commonly accepted best criteria to decide upon the threshold to cut dendrograms in unsupervised learning. There is no known information (like labels in supervised learning) to guide such decisions. However, we figured out a perfect solution for determining the threshold by combining unsupervised learning with supervised learning in our situation. We consider many candidate thresholds for cutting the dendrograms. Each candidate value corresponds to one grouping rule and leads to a fitted Elastic Net model based on the grouped genes, creating a one-to-one correspondence between a threshold value and the performance of the final prediction of cell phenotypes. We use $10$-fold cross-validation to compare the prediction performance (e.g. AUC statistics for binary outcomes and MSE statistics for continuous outcomes) of the Elastic Net models associated with those candidate thresholds. The threshold associated with the winner prediction model is treated as the best one, and we use that threshold to obtain the final gene grouping. At last, we fit the Elastic Net model using the grouped genes to get the final prediction model. In short, we consider the tree trimming threshold as equivalent to a parameter of Elastic net models and use cross-validation to select its best value.
We summarize our proposed method in Algorithm~\ref{alg2}. 

\begin{algorithm}
	\caption{Integration of the Grouping Method with the Elastic Net Method}
	\label{alg2}
	\begin{algorithmic}
		\Input \\
		(1) Gene expression level: $\boldsymbol{x}$, a $n \times p$ matrix\\ 
		(2) Cell phenotype: $\boldsymbol{y} = (y_1, \ldots, y_n)$ , a binary vector of length $n$ (e.g. indicator of a cell subtype)\\
		Note: $n$ is the number of cells and $p$ is the number of genes. \\
		
		\Algorithm
		\State (1) Pre-group the genes into smaller subsets using Algorithm~\ref{alg1}. 
		\State (2) In each subset of genes, apply the hierarchical clustering algorithm to further cluster genes, and build the dendrogram. We use (1-correlation) as dissimilarity and use the average link.
		\State (3) Randomly split cells into $10$ folds containing equal number of cells.
		\For{ $c \in \{10^{-1}, 5\times10^{-2},10^{-2}, 5\times10^{-3},10^{-3}, 5\times10^{-4},10^{-4}, 5\times10^{-5},10^{-5}, 5\times10^{-6},10^{-6}\}$}
		\State (4.1) Cut all dendrograms using dissimilarity threshold $c$, and group all genes in the same branch after cutting. Calculate the common pattern $\boldsymbol{z}$ of the gene groups. 
		\State (4.2) For $i=1, \ldots, 10$, fit the Elastic Net model to data excluding cells in the $i$-th fold, and predict phenotypes of cells in the $i$-th fold $\hat{q}_j=\Pr(y_j=1)$.
		\State (4.3) Pool $\hat{q}_j$ obtained from $10$ folds to form cross-validation predictions on full data.
		\State (4.4) Calculate AUC statistics from $y_j$'s and $\hat{q}_j$'s.
		\EndFor
		\State (5) Obtain final gene-grouping results by cutting all dendrograms with the threshold value $c$ corresponding to the largest AUC.
		\State (6) Fit the Elastic Net model based on final gene-grouping results.
		\Output \\
		The Elastic Net model based on final gene-grouping results.
	\end{algorithmic}
\end{algorithm}

\section*{Simulation Study}
This section conducts simulation studies to compare the integration of the grouping and the Elastic Net methods (algorithm 2) and the standard Elastic Net method. The results show that our novel integration method improves the accuracy of prediction of cell phenotypes and the ability to select true feature genes. Our comparison is based on the recently published scRNA-seq benchmark datasets used to compare the performance of DE methods\cite{Soneson:2018df}. We use the same simulation strategy mentioned in their paper, i.e., assuming a data generation model with model parameters learned from real data. We revised some details in data simulation to make conditions closer to the real-world settings.
      
\subsection*{Benchmark datasets and simulation design} \label{sec:benchmark}
Soneson and Robinson \cite{Soneson:2018df} developed Conquer, a collection of consistently processed scRNA-seq benchmark datasets. They conducted a simulation study based on $6$ real datasets, i.e. GSE74596 \cite{Engel:2016gc},  GSE63818-GPL16791 \cite{Guo:2015ff},  GSE60749-GPL13112 \cite{Kumar:2014kv},  GSE48968-GPL13112 \cite{Shalek:2014eya},   GSE45719 \cite{Deng:2014hd}, and EMTAB2805 \cite{Buettner:2015hp}. In our work, we use the same genomic datasets to evaluate the performance of prediction analysis. We consider the binary cell phenotype as the outcome, which is defined the same as given the Supplementary Table 1 of Soneson and Robinson's paper \cite{Soneson:2018df}, which was used to compare the performance of $36$ DE methods. 
 
\hspace{-0.2in} \textbf{The simulation model} In the Conquer study, they generated synthetic genomic datasets to evaluate the performance of DE methods, whose reliability depends on the agreement between the assumed distributions and the underlying truth. Comparatively, our goal is to improve prediction and ability in feature gene selection. So we use those real-world genomic data, particularly the $6$ full-length data, to retain the natural complex structure among genes and only simulate the cell phenotypes (i.e. binary group memberships) from a logistic regression model
\begin{align}
 \label{f:sim1}  y_j &\sim \mbox{Bernoulli}(q_j)\;\; \mbox{ and } \;\;q_j = \mbox{logit}^{-1} \left( \beta_{0} + \sum_{i=1}^p \beta_i x_{ji} \right) 
\end{align}
where $i$ is the index of genes, $j$ is the index of cells, $y_j$ is the group membership indicator of the $j$-th cell, $x_{ji}$ is the expression level (transcripts per million) of the $i$-th gene of the $j$-th cell, and $\beta_i$ is the effect size (or coefficient) of the $i$-th gene with $i = 1, 2, \cdots, p$ and $j = 1, 2, \cdots, n$.


\hspace{-0.2in} \textbf{The coefficients and jittering} To assign the values of coefficients $\beta_i$'s in the data-generating model~(\ref{f:sim1}), we fit an Elastic Net model to each of the $6$ real datasets. The genes with non-zero fitted coefficients (i.e. $\beta_i \neq 0$) are considered as feature genes in the data-generating model~(\ref{f:sim1}). When the feature genes are not highly correlated with other genes, our proposed method will not group them with any other genes, therefore our method becomes identical to the standard (ungrouped) method. To make the simulation comparison meaningful, when fitting the Elastic Net models, we force genes to have $0$ coefficients if their correlations with any other genes are less than $0.9$. In this way, we focus the method comparisons on situations where feature genes are highly correlated with other genes. 

Those $6$ fitted Elastic Net models based on the benchmark datasets are the data generating models. We call them the `blueprint models'. To increase the heterogeneity of simulated datasets, we add Gaussian noises to the coefficients of each blueprint model to generate $100$ different jittered data-generation models. Based on each jittered data-generation model and its corresponding real genomic data, we simulate one copy of the phenotype, i.e. binary cell type membership, for every cell in the corresponding genomic data. 
In summary, we first generate one blueprint model from each benchmark dataset, add jitters to create $100$ data-generation models, and finally obtain $100$ copies of simulated cell phenotypes, which results in $600$ simulated datasets for methods comparison.  

\hspace{-0.2in} \textbf{Data analysis and evaluation criteria}
We analyze those $600$ simulated datasets using the standard Elastic Net model and the integration of our proposed grouping method with the Elastic Net. We denote $\hat{\beta}_i$ and $\hat{q}_j$ as the estimated values for parameter $\beta_i$ and $q_j$ respectively,  $(\mbox{for } i = 1, \ldots, p \mbox{ and } j = 1,  2, \ldots, n)$, and calculate the following four criteria to compare the two methods for all simulated data.  
\begin{align}
\label{f:crit1} \mbox{MSE} &=  \frac{\sum_{j=1}^n (q_{j} - \hat{q}_{j})^2}{n}, \\
\label{f:crit2} \mbox{Precision} &= \frac{\sum_{i=1}^p I(\beta_{i} \neq 0)I(\hat{\beta}_{i} \neq 0) }{ \sum_{i=1}^p I(\hat{\beta}_{i} \neq 0) }, \\
\label{f:crit3} \mbox{Recall} &= \frac{\sum_{i=1}^p I(\beta_{i} \neq 0)I(\hat{\beta}_{i} \neq 0) }{ \sum_{i=1}^p I(\beta_{i} \neq 0) }, \\
\label{f:crit4} \mbox{F1} &= 2 (\mbox{Precision}^{-1} + \mbox{Recall}^{-1})^{-1}.
\end{align}

\subsection*{Comparison of prediction performance}
To evaluate the prediction performance based on real data, the most popular criteria are the misclassification rate ( i.e. 1-accuracy) and the AUC statistics, which can be derived from the estimated parameter, $\hat{q}_j$, and the observed data $y_j$ (a realization of a Bernoulli random variable with parameter $q_j$).  However, in this simulation, since the generating probability of the group membership, $q_j$, are known, we use MSE~(\ref{f:crit1}), a more direct and precise measurement, to evaluate the prediction performance. We calculate the pairwise difference of MSE obtained from the integrated method and the standard Elastic Net model for each simulated dataset. A negative difference indicates that the integration method outperforms the Elastic Net model.  

\begin{figure}[h!btp]
\begin{center}
\vspace{-0.1in}
\includegraphics[scale=0.15]{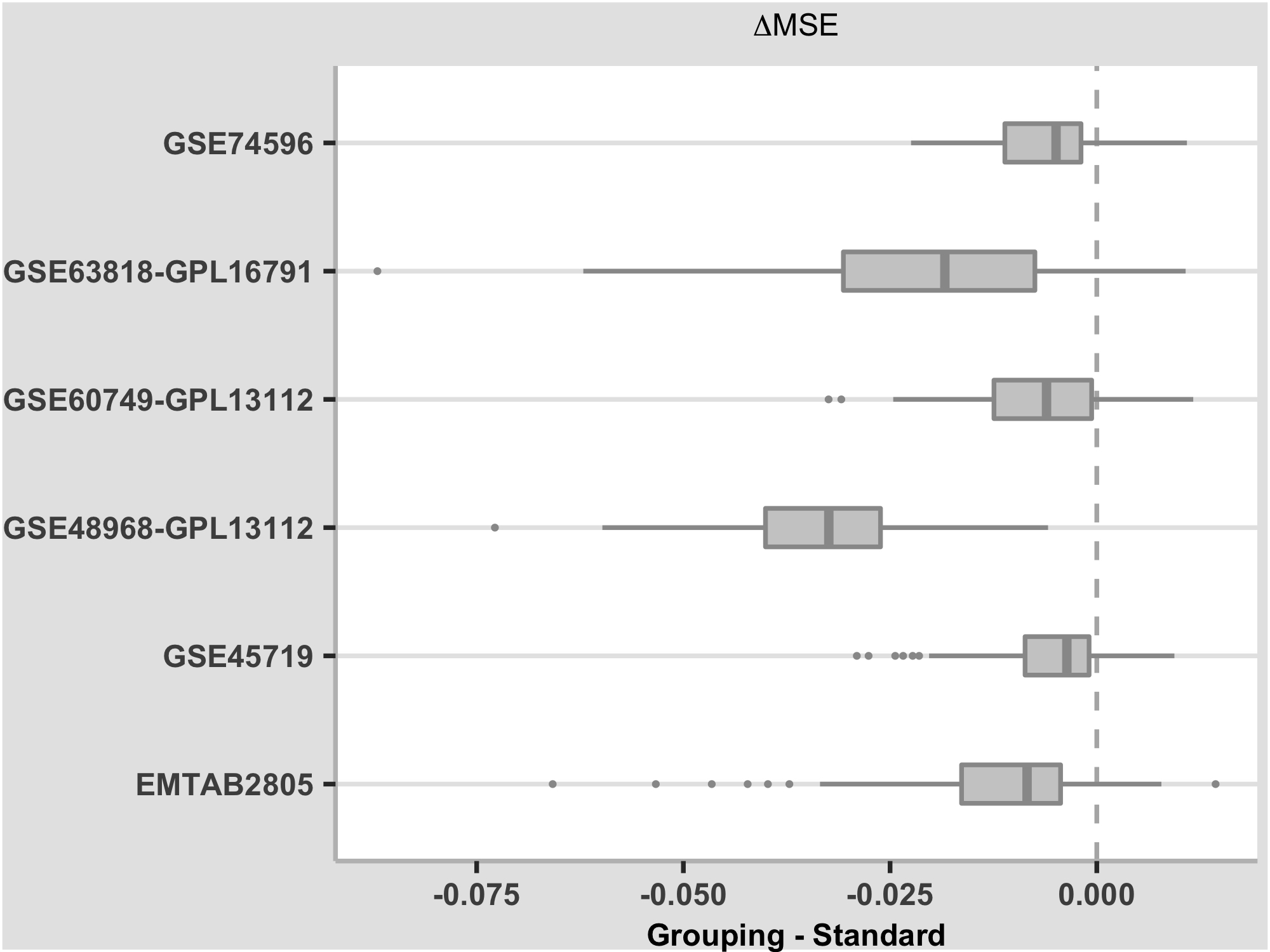}
\caption{Boxplots of differences in the mean squared prediction error between the integration method (grouping + Elastic Net) and the standard method. Each row represents the results of $100$ simulated datasets based on one real scRNA-seq benchmark data.} \label{fig:MSE}
\end{center}
\end{figure}
Figure \ref{fig:MSE} visualizes the $600$ differences in MSE calculated from our simulation studies. Each box represents the results of $100$ datasets simulated from one benchmark scRNA-seq data. The name of each benchmark data is labelled on the left-hand side of the boxes. The majority of the values in the boxplots are less than $0$, which indicates that our method enhances the prediction performance of the standard Elastic Net model. To investigate the significance of those differences visualized in the boxplots, we conducted paired Wilcoxon signed-rank test to compare those MSE's and the resulting p-values are all less than $10^{-10}$. In summary, the simulation results demonstrate that the prediction performance of the Elastic Net model can be significantly improved by integrating with our novel grouping strategy.

\subsection*{Comparison of performance on feature gene selection}
Selecting feature genes can be considered a series of binary decisions for all the genes. In the machine learning community, the quantities~(\ref{f:crit2})-~(\ref{f:crit4}) are the most popular criteria to evaluate the performance of feature selection. Precision~(\ref{f:crit2}) is the fraction of the true feature gene among the selected ones, which is equivalent to $1 - $False Discovery Rate. Recall~(\ref{f:crit3}) is the fraction of the selected true genes among the true feature genes, which is also called sensitivity in the statistical community. When evaluating the performance of methods, precision and recall need to be considered together as a pair. These two criteria may rank the performance of the methods differently, while the F1 score~(\ref{f:crit4}) is the harmonic mean of precision and recall, which carries information of both statistics.

\begin{figure}[h!btp]
\begin{center}
\vspace{-0.1in}
\includegraphics[scale=0.15]{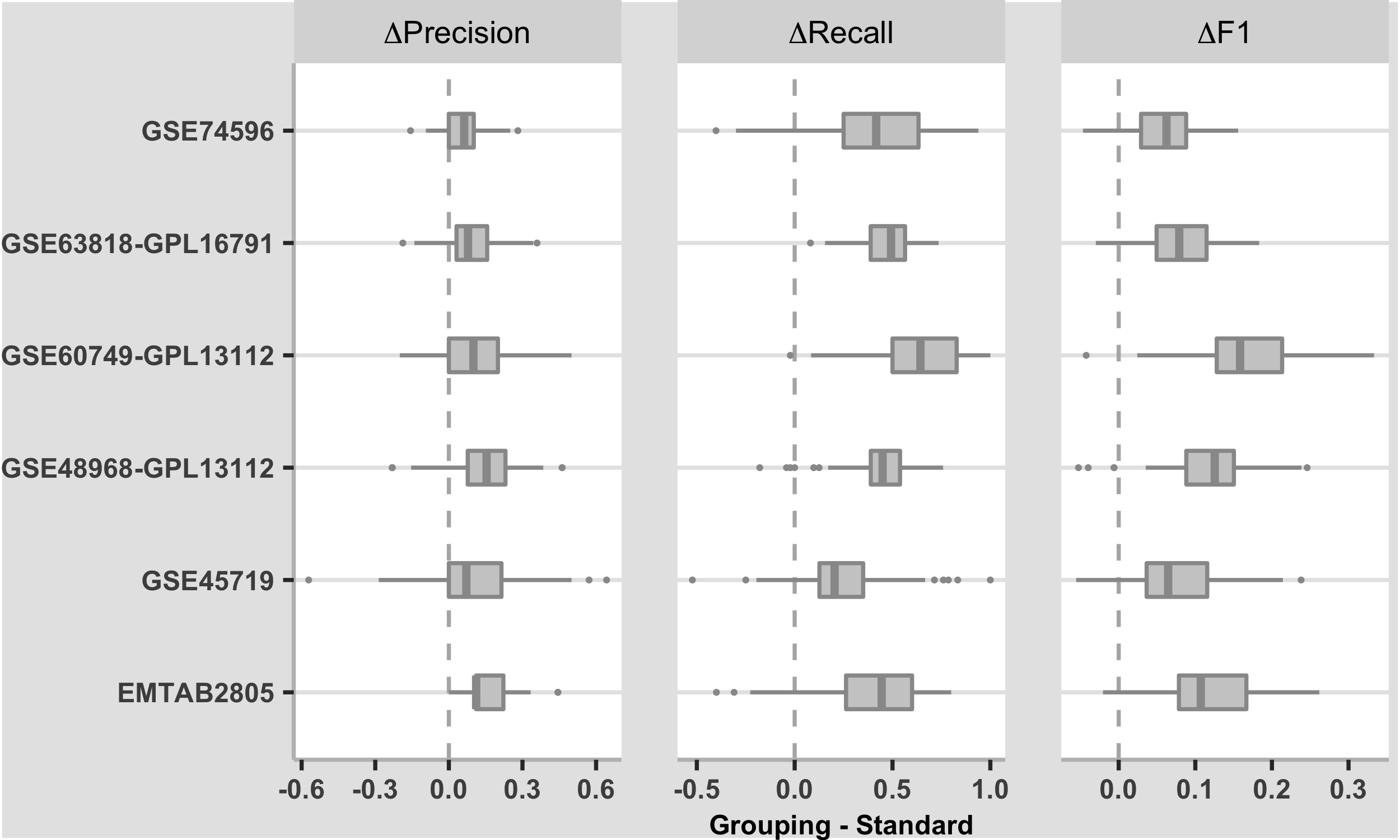}
\caption{Boxplots of paired differences between the integration method (grouping + Elastic Net) and the standard method in three feature selection criteria. Precision in the left panel, Recall in the middle panel, and F1 score in the right panel. Each row shows the results of simulation based on one benchmark data.} \label{fig:PR}
\end{center}
\end{figure}

Figure \ref{fig:PR} visualizes the pairwise differences (the integrated one minus the standard one) of the precision (left panel), recall (middle panel) and F1 scores (right panel) calculated from each simulated data. Each row represents pairwise differences from $100$ simulated datasets based on one benchmark scRNA-seq data labelled on the left-hand side of boxes. Positive values of those differences indicate that the integration of our method with the Elastic Net outperforms the standard Elastic Net. So, those boxes with the majority of the differences greater than $0$ indicate that the integrated method outperforms the standard Elastic Net model in gene selection. To investigate the significance of those differences, we conducted paired Wilcoxon signed-rank tests, and the resulting p-values are less than $10^{-5}$ for the precision differences and $10^{-14}$ for the recall differences and F1 score differences. In summary, the simulation result shows that the performance of feature gene selection of the Elastic Net is significantly improved by integrating with our novel grouping strategy.

\section*{Discussion and Conclusion}
The prediction analysis discussed in this work should not be confused with the other common type of analysis, Differential Expression (DE) analysis. Even though both of them involve discovering genes, the two kinds have very different objectives. The objective of DE is to identify genes significantly differentially expressed between two conditions, such as two cell phenotypes, in scRNA-seq analysis. DE analysis tests genes one by one, and its conclusions are based on the resulting p-values. Prediction analysis discovers a parsimonious list of genes that `work together' to predict the cell phenotypes. Its focus is on the model's prediction performance (but not p-values) built based on the selected gene lists. The strong correlations between genes are helpful in DE analyses since they enable individual tests of a single genomic marker to borrow information from each other \cite{Xu:2019gi}. However, in prediction analysis, strong correlations between predictors (if not properly addressed) can introduce inconsistent results in prediction and feature selection. Therefore, properly incorporating the strong correlations among genes is critical for improving the model performance in prediction.

Our proposed grouping method is designed to build a stable and accurate prediction model at the existence of the strong correlations between genes. It has multiple advantages. First, we use a single variable to replace a group of highly correlated genes, which avoids the multi-collinearity problems \cite{Farrar:1967bb} caused by highly correlated genes. Therefore, we have a more stable prediction model. Second, in feature selection, highly correlated predictors always compete for importance, which leads to being reported as less important than they should be. By using a single predictor to represent the group, we avoid such competition. Third, our method presents the whole group of highly correlated genes as feature genes instead of just picking one gene (as in the standard methods). As shown in simulation studies, to predict new cells' phenotypes, using the common pattern of the gene group is more robust than using a single gene in the group. Most importantly, reporting the whole group of highly correlated genes as feature genes allows researchers to identify causal genes in follow-up studies. The causal genes are critical in real-world applications. In a hypercritical example, we assume that genes A and B are highly correlated, and a higher expression level of gene-A causes both a higher expression level of gene-B and disease. When it is developed targeted at reducing the expression level of gene-B, the drug will not have efficacy for the disease. Instead, the drug should be designed to decrease gene-A expression to treat the disease. 

Our grouping method is an example of integrating the supervised machine learning methods with unsupervised ones. The objective of our research is to improve the prediction, which is a supervised machine learning problem. To address the particular challenge in genomic studies, we propose the grouping strategy, which is unsupervised in nature. During this process, there is no consistently accepted best way to decide the number of clusters or the threshold for cutting dendrogram since there is no way to validate the correctness of the clustering rule. We proposed a good solution in our special situation. Each clustering result corresponds to a grouping rule for the genes, leading to a different prediction. This one-to-one correspondence between the clustering rule and the prediction performance enables us to utilize the prediction results to determine the number of clusters (or, say, the threshold for cutting dendrogram of hierarchical clustering results). We use cross-validation to choose the best value. Finally, the selected best clustering rule is used to group genes for building the final prediction model. Such integration of supervised and unsupervised learning decides the best number of clusters and provides better prediction performance.

In addition, our method has more general applications. First, our method can be integrated with other supervised learners but not limited to the Elastic Net model. Second, our method can be applied to the analysis of different types of datasets (not only the scRNA-seq data) with two characteristics: (1) predictors are highly correlated and (2) the correlation structures between predictors are not consistent in the training and the test data.  

In summary, we propose a grouping algorithm to handle highly correlated genes in prediction analysis. We showed that the method significantly improves cell phenotype classification and feature gene selection in simulation studies based on multiple published benchmark scRNA-seq datasets. Most importantly, our method reports important gene groups instead of individual genes as final predictors, leading to a more robust and powerful prediction model and allowing other researchers to identify causal genes (within these groups) in downstream analysis.

\section*{Availability of data and materials}
The datasets in this work are the same as those in the reference paper \cite{Soneson:2018df}, available on Drs Soneson and Robinson's website \url{http://imlspenticton.uzh.ch/robinson_lab/conquer_de_comparison/}. 

The proposed method is implemented into an R package. We are finalizing the documentation of the package and will submit it to CRAN soon.

\section*{Acknowledgements}
The authors thank Dr. Charlotte Soneson and Prof. Mark Robinson (University of Zurich) for their helpful discussion about the benchmark datasets and their simulation strategy for comparing DE methods of scRNA-seq data.

\section*{Funding}
This work was supported by the Natural Sciences and Engineering Research Council Discovery Grants (XZ, LX, ML), the Canada Research Chair (XZ), and Compute Canada RAC (XZ, LX). This research was enabled in part by support provided by WestGrid (www.westgrid.ca) and Compute Canada (www.computecanada.ca).

\section*{Author contributions statement}
XZ and LX contributed to the study concept and design, developed the algorithm, implemented the algorithm in R code, and prepared the first draft manuscript. SJ and KM conducted the simulation studies. XZ and ML supervised the project’s progress. All authors contributed to revise the draft manuscript, read and approve the final manuscript.

%
%

\begin{thebibliography}{9}

\bibitem{Pliner2019}
Pliner, H. A., Shendure, J.,  Trapnell, C. (2019) Supervised classification enables rapid annotation of cell atlases. \textit{Nature Methods}, 16, 983–986 (2019).  

\bibitem{Angelidis2019fw}
Angelidis, I., Simon, L.~M., Fernandez, I.~E., Strunz, M., Mayr, C.~H., Greiffo, F.~R., Tsitsiridis, G., Ansari, M., Graf, E., Strom, T.-M., et~al (2019) An atlas of the aging lung mapped by single cell transcriptomics and deep tissue proteomics. \textit{Nature communications}, 10(1):963--17.

\bibitem{Aran:2017do}
Aran, D., Hu, Z., and Butte, A.~J. (2017) xCell: digitally portraying the tissue cellular heterogeneity landscape. \textit{Genome Biology}, 18(1):220--14.

\bibitem{Buettner:2015hp}
Buettner, F., Natarajan, K.~N., Casale, F.~P., Proserpio, V., Scialdone, A., Theis, F.~J., Teichmann, S.~A., Marioni, J.~C., and Stegle, O. (2015) Computational analysis of cell-to-cell heterogeneity in single-cell RNA-sequencing data reveals hidden subpopulations of cells. \textit{Nature biotechnology}, 33(2):155--160.

\bibitem{Deng:2014hd}
Deng, Q., Ramsk{\"o}ld, D., Reinius, B., and Sandberg, R. (2014) Single-cell RNA-seq reveals dynamic, random monoallelic gene expression in mammalian cells. \textit{Science (New York, NY)}, 343(6167):193--196.

\bibitem{Dixit:2016fe}
Dixit, A., Parnas, O., Li, B., Chen, J., Fulco, C.~P., Jerby-Arnon, L.,
Marjanovic, N.~D., Dionne, D., Burks, T., Raychowdhury, R., et~al (2016) Perturb-Seq: Dissecting Molecular Circuits with Scalable Single-Cell
RNA Profiling of Pooled Genetic Screens.
\textit{Cell}, 167(7):1853--1866.e17.

\bibitem{Engel:2016gc}
Engel, I., Seumois, G., Chavez, L., Samaniego-Castruita, D., White, B., Chawla, A., Mock, D., Vijayanand, P., and Kronenberg, M. (2016) Innate-like functions of natural killer T cell subsets result from highly divergent gene programs. \textit{Nature immunology}, 17(6):728--739.

\bibitem{Farrar:1967bb}
Farrar, D.~E. and Glauber, R.~R. (1967) Multicollinearity in Regression Analysis: The Problem Revisited. \textit{The Review of Economics and Statistics}, 49(1):92.

\bibitem{Gawad:2016eq}
Gawad, C., Koh, W., and Quake, S.~R. (2016) Single-cell genome sequencing: current state of the science. \textit{Nature Reviews Genetics}, {\bf 17}(3):175--188.

\bibitem{Geeleher:2014fa}
Geeleher, P., Cox, N.~J., and Huang, R. (2014) Clinical drug response can be predicted using baseline gene expression levels and in vitro drug sensitivity in cell lines.
\textit{Genome Biology}, 15(3):R47.

\bibitem{Guo:2015ff}
Guo, F., Yan, L., Guo, H., Li, L., Hu, B., Zhao, Y., Yong, J., Hu, Y., Wang, X., Wei, Y., et~al (2015) The Transcriptome and DNA Methylome Landscapes of Human Primordial Germ Cells. \textit{Cell}, 161(6):1437--1452.

\bibitem{Gusev:2016ey}
Gusev, A., Ko, A., Shi, H., Bhatia, G., Chung, W., Penninx, B. W. J.~H.,
Jansen, R., de~Geus, E. J.~C., Boomsma, D.~I., Wright, F.~A., et~al (2016) Integrative approaches for large-scale transcriptome-wide association studies. \textit{Nature Publishing Group}, 48(3):245--252.

\bibitem{Herberg:2016fi}
Herberg, J.~A., Kaforou, M., Wright, V.~J., Shailes, H., Eleftherohorinou, H.,
Hoggart, C.~J., Cebey-L{\'o}pez, M., Carter, M.~J., Janes, V.~A., Gormley,
S., et~al (2016) Diagnostic Test Accuracy of a 2-Transcript Host RNA Signature for
Discriminating Bacterial vs Viral Infection in Febrile Children.
\textit{JAMA}, 316(8):835--845.

\bibitem{Kiselev:2019bk}
Kiselev, V.~Y., Andrews, T.~S., and Hemberg, M. (2019) Challenges in unsupervised clustering of single-cell RNA-seq data.
\textit{Nat Rev Genet}, 20(5):273--282.

\bibitem{Kumar:2014kv}
Kumar, R.~M., Cahan, P., Shalek, A.~K., Satija, R., DaleyKeyser, A., Li, H.,
Zhang, J., Pardee, K., Gennert, D., Trombetta, J.~J., et~al (2014) Deconstructing transcriptional heterogeneity in pluripotent stem
cells. \textit{Nature}, 516(7529):56--61.

\bibitem{Nawy:2014wa}
Nawy, T. (2014) Single-cell sequencing. \textit{Nature methods}, 11(1):18--18.

\bibitem{Newman:2019ko}
Newman, A.~M., Steen, C.~B., Liu, C.~L., Gentles, A.~J., Chaudhuri, A.~A.,
Scherer, F., Khodadoust, M.~S., Esfahani, M.~S., Luca, B.~A., Steiner, D.,
et~al (2019) Determining cell type abundance and expression from bulk tissues
with digital cytometry. \textit{Nature biotechnology}, 37(7):773--782.

\bibitem{Richardson:2020co}
Richardson, T.~G., Hemani, G., Gaunt, T.~R., Relton, C.~L., and Davey~Smith, G.
(2020) A transcriptome-wide Mendelian randomization study to uncover tissue-dependent regulatory mechanisms across the human phenome. \textit{Nature communications}, 11(1):185--11.

\bibitem{Shalek:2014eya}
Shalek, A.~K., Satija, R., Shuga, J., Trombetta, J.~J., Gennert, D., Lu, D.,
Chen, P., Gertner, R.~S., Gaublomme, J.~T., Yosef, N., et~al (2014) Single-cell RNA-seq reveals dynamic paracrine control of cellular variation.
\textit{Nature}, 510(7505):363--369.

\bibitem{Soneson:2018df}
Soneson, C. and Robinson, M.~D. (2018) Bias, robustness and scalability in single-cell differential expression analysis. \textit{ Nature methods}, 15(4):255--261.

\bibitem{Wagner:2019gz}
Wagner, J., Rapsomaniki, M.~A., Chevrier, S., Anzeneder, T., Langwieder, C., Dykgers, A., Rees, M., Ramaswamy, A., Muenst, S., Soysal, S.~D., et~al (2019) A Single-Cell Atlas of the Tumor and Immune Ecosystem of Human Breast Cancer.
\textit{Cell}, 177(5):1330--1345.e18.

\bibitem{Xing:2019ji}
Xing, L., Lesperance, M.~L., and Zhang, X. (2019) Simultaneous prediction of multiple outcomes using revised stacking algorithms. \textit{Bioinformatics (Oxford, England)}, 36(1):65--72.

\bibitem{Xu:2019gi}
Xu, Y., Xing, L., Su, J., Zhang, X., and Qiu, W. (2019) Model-based clustering for identifying disease-associated SNPs in case-control genome-wide association studies. \textit{Scientific reports}, 9(1):13686--10.

\bibitem{Zou:2005ex}
Zou, H. and Hastie, T. (2005) Regularization and variable selection via the elastic net. \textit{Journal of the Royal Statistical Society. Series B. Statistical Methodology}, 67(2):301--320.
\end{thebibliography}








\end{document}